\documentclass[prb,twocolumn,aps,showpacs,fixfloats]{revtex4}
\usepackage{graphicx}
\usepackage{bm}
\usepackage{amsmath,amssymb}
\usepackage{subfigure}
\usepackage{float}
\usepackage{latexsym}
\usepackage{epstopdf}
\usepackage{color}
\usepackage{enumerate}

\begin{document}
\newcommand{\s}{\scriptscriptstyle}
\newcommand{\uu}{\uparrow \uparrow}
\newcommand{\ud}{\uparrow \downarrow}
\newcommand{\du}{\downarrow \uparrow}
\newcommand{\dd}{\downarrow \downarrow}
\newcommand{\ket}[1] { \left|{#1}\right> }
\newcommand{\bra}[1] { \left<{#1}\right| }
\newcommand{\bracket}[2] {\left< \left. {#1} \right| {#2} \right>}
\newcommand{\vc}[1] {\ensuremath {\bm {#1}}}
\newcommand{\tr}{\text{Tr}}

\title{Organic magnetoresistance under resonant ac drive}

\author{R. C. Roundy  and M. E. Raikh} \affiliation{Department of Physics and
Astronomy, University of Utah, Salt Lake City, UT 84112}

\begin{abstract}
We study the spin dynamics of an electron-hole polaron pair in a random
hyperfine magnetic field and an external field, $B_{\s 0}$, under a resonant
drive with frequency $\omega_{\s 0}=\gamma B_{\s 0}$.  The fact that the pair
decays by recombination exclusively from a singlet configuration, $S$, in which
the spins of the pair-partners are entangled, makes this dynamics highly
nontrivial. Namely, as the amplitude, $B_{\s 1}$, of the driving field grows,
mixing  all of the triplet components, the long-living modes do not disappear,
but evolve from $T_{\s +}$, $T_{\s -}$ into $\frac{1}{2}\left(T_{\s +}\pm
\sqrt{2}T_{\s 0} +T_{\s -}\right)$.  Upon further increase of $B_{\s 1}$, the
lifetime of the $S$-mode is cut in half, while the $T_{\s 0}$-mode transforms
into an antisymmetric combination $\frac{1}{\sqrt 2}\left(T_{\s +}-T_{\s
-}\right)$ and acquires a long lifetime, in full analogy to the superradiant
and subradiant modes in the Dicke effect.  Peculiar spin dynamics translates
into a peculiar dependence of the current through an organic device on $B_{\s
1}$. In particular, at small $B_{\s 1}$, the radiation-induced correction to
the current is {\em linear} in $B_{\s 1}$.
\end{abstract}

\maketitle

\noindent
{\em Introduction.}
Experimental finding that the intensity of room-temperature
exciton luminescence in  anthracene  crystal changes
by several percent in a weak magnetic field $B\sim 0.1$~T
was reported more than four decades ago\cite{luminescence}.
Such a small scale of $B$ is set by the magnitude of zero-field splitting which controls the spin states of a pair of annihilating carriers forming an exciton. Organic magnetoresistance
(OMAR) is an effect of a similar physical origin,
where the external magnetic field causes a change
of current through an organic layer\cite{Markus0,Markus1,Markus2}
by affecting the rate  of spin-dependent processes, either
recombination\cite{Prigodin,Gillin} or bipolaron formation\cite{Bobbert1,BobbertStochastic}. It is commonly accepted
that, in OMAR, the scale of $B$ is set by random hyperfine fields with
rms $b_{\s 0} \sim 10$~mT. This fundamental origin of OMAR explains why
the effect itself is robust, while its magnitude and even the sign
are sensitive to technological details\cite{XuWu,Valy0,Blum,Valy1}.

\begin{figure}[t]
\includegraphics[width=77mm, clip]{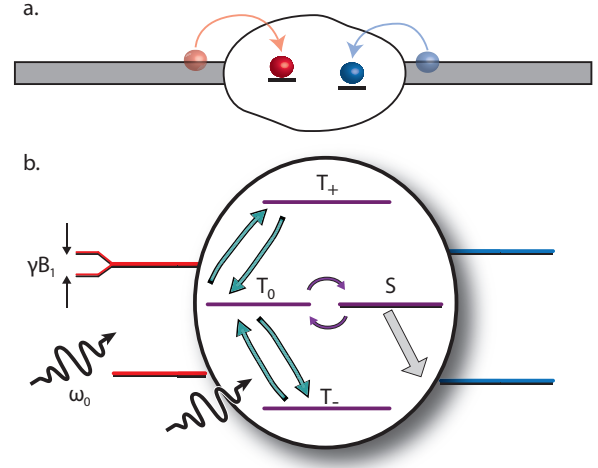}
\caption{(Color online). (a) Current passage through a bipolar device involves recombination of electron (red) and hole (blue) which occupy the neighboring sites; (b) Example of a pair in which electron is on-resonance and hole is off-resonance. The bubble
illustrates the efficient mixing of the triplet components by the ac field, which, in turn,
affects the crossing rate $T_{\s 0}\leftrightarrows S$. The gray arrow indicates that recombination occurs exclusively from $S$.  }
\label{irradiation}
\end{figure}

\begin{figure}[t]
\includegraphics[width=77mm, clip]{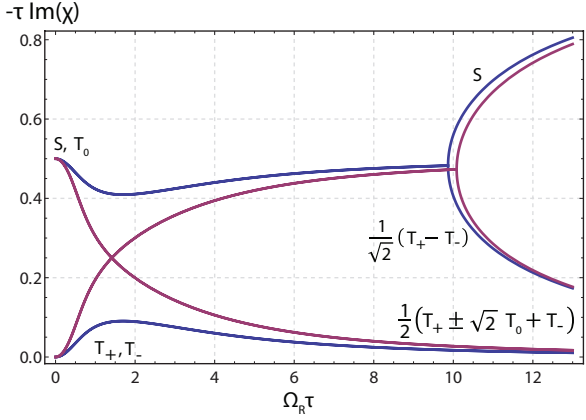}
\caption{(Color online). The evolution of dimensionless decay rates of different modes with amplitude of the ac drive is plotted from Eq. (\ref{quartic-recombination})
for two sets of parameters $\left(\delta\tau, \delta_{\s 0}\tau\right)$: blue $\left(2.5, 2\right)$; purple $\left(2, 2.5\right)$. The content of the quasimodes evolves from
$T_{\s +}, T_{\s -}$ and linear combinations of $S$, $T_{\s 0}$ at weak drive into the combinations, $\frac{1}{2}\left(T_{\s +}\pm \sqrt{2}T_{\s 0}
+T_{\s -}\right)$,
one superradiant mode, $S$, and one subradiant mode, $\frac{1}{\sqrt 2}\left(T_{\s +}-T_{\s -}\right)$.  }
\label{im-chi}
\end{figure}

To capture the fundamental nature of OMAR quantitatively, it is
sufficient  to adopt the simplest assumption\cite{Bobbert1,Flatte1, we}
that bipolaron formation or recombination proceed only when the
pair-partners are in the singlet state, $S$. With equal probabilities of
all initial states, the recombination time of a pair
is determined by the degree of admixture of the singlet to
three other spin eigenstates caused by the hyperfine field.
For external field $B\sim b_0$ the current response, $I(B)$,
is governed by {\em blocking} configurations\cite{Bobbert1} in
which hyperfine fields ``conspire'' to protect the pair from crossing into $S$
after its creation. As the field increases and exceeds $b_{\s 0}$,  these long-living states evolve into
$T_{+}$ and $T_{-}$ components of a triplet, and the current saturates.

A recent experiment, Ref. \onlinecite{Baker}, has demonstrated
that saturated OMAR exhibits a lively response to the
external resonant ac drive at frequency $\omega_{\s 0} =\gamma B_{\s 0}$, where $\gamma$ is the gyromagnetic ratio.
The experiment was performed on a bipolar organic-semiconductor diode
placed on the top of a conducting stripline through which the ac current
was passed.   To the first approximation, this fascinating finding can be
accounted for by considering the ac field
as a mixing agent, which tends to scramble all three triplet states  and, thus, to limit the trapping ability of $T_{\s +}$, $T_{\s -}$, see Fig. \ref{irradiation}.
In this way, the ac field tends to change the current towards its
value at zero magnetic field, which is what was observed in Ref. \onlinecite{Baker}. From the above picture one would expect that the radiation-induced change of
current, $\delta I$, is due to the change of the recombination rate, which, in turn, is proportional
to $B_{\s 1}^2$, i.e. to the power of the driving field.

In the present paper we demonstrate that the dependence of $\delta I$ on
$B_{\s 1}$ is much more intricate. In particular, it is {\em linear} for
weak $B_{\s 1}$. This effect stems from pairs in which one
of the partners is on-resonance, see Fig. \ref{irradiation}.
It appears that for these particular pairs the radiation-induced suppression of trapping by  $T_{\s +}$ and $T_{\s -}$ is especially efficient.
However, such pairs determine $\delta I(B_{\s 1})$ only for weak
driving fields, namely, for  fields in which the
nutation frequency is much smaller than $\gamma b_{\s 0}$.
As we demonstrate below, a very nontrivial physics unfolds
for  higher $B_{\s 1}$.
Quite unexpectedly, a new {\em long-living mode},
$\frac{1}{\sqrt{2}}\left(T_{\s +}-T_{\s -}\right)$, emerges
 in strong enough driving fields, see Fig. \ref{im-chi}.
This mode, in which both pair-partners are on resonance, is fully analogous to subradiant state in the Dicke effect\cite{Dicke}. Trapping by this
state also yields a linear correction to the current, but with {\em opposite slope}.

\begin{figure}[t]
\includegraphics[width=77mm, clip]{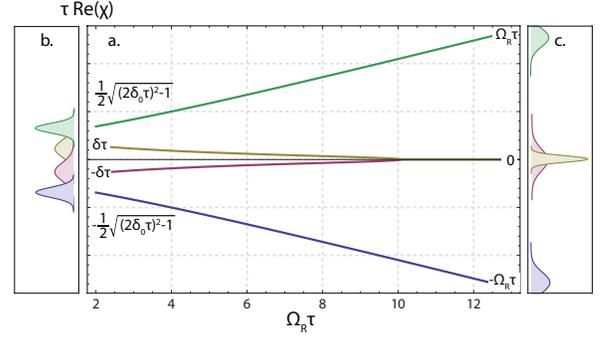}
\caption{(Color online). (a) The evolution of quasienergies with amplitude of the driving field is plotted from Eq. (\ref{quartic-recombination}) for parameters $(\delta\tau, \delta_{\s 0}\tau)= (2, 2.5)$.  Quasienergies evolve from $\pm \delta$, $\pm \frac{1}{2} \sqrt{ (2 \delta_{\s 0} \tau)^2 -1}$ to $0, \pm \Omega_{\s R}$. At small $\Omega_{\s R}$, the quasienergies are well resolved (b). Merging of two quasienergies at large $\Omega_{\s R}$ is accompanied by       splitting of their widths (c), which is a manifestation of the Dicke physics.}
\label{re-chi}
\end{figure}

\noindent
{\em Driven spin-pair without recombination.}
To highlight the physics, we first neglect recombination.
The Hamiltonian of the driven pair has a form
\begin{equation}
\label{Hamiltonian}
\widehat{H} = \omega_{\s e}S^z_{\s e}
+\omega_{\s h}S^z_{\s h}
+ 2 \Omega_{\s R} \left(S^x_{\s e} + S^x_{\s h} \right)
\cos \omega_{\s 0} t,
\end{equation}
where $\omega_{\s e,h}=\omega_{\s 0}+\delta_{\s e,h}$, $\Omega_{\s R}=\gamma B_{\s 1}$ is the Rabi frequency, and $\delta_{\s e,h}$ are the $z$-components
of the hyperfine fields acting on the electron and hole, respectively, i.e. the detunings of the
pair-partners from  the resonance. By retaining only $z$-components, we assumed that $B_{\s 0} \gg b_{\s 0}$.
We will also assume that $\gamma B_{\s 0} \gg \Omega_{\s R}$, which allows us to employ  the rotating wave approximation.
In the rotating frame, the amplitudes of $T_{\s +}$, $T_{\s -}$, $T_{\s 0}$, and $S$-components of the wave function
are related as
\begin{align}
\label{driven-system}
(\chi - \delta)A_{\s T_{\s -}} = \frac{\Omega_{\s R}}{\sqrt{2}} A_{\s T_{\s 0}}, \quad
(\chi + \delta)A_{\s T_{\s +}} = \frac{\Omega_{\s R}}{\sqrt{2}} A_{\s T_{\s 0}},  \\
\!\chi A_{\s S} =-
\delta_{\s 0} A_{\s T_{\s 0}}, \;
\!\chi A_{\s T_{\s 0}} \!= -
\delta_{\s 0} A_{\s S}
    + \!\frac{\Omega_{\s R}}{\sqrt{2}} \left(
     A_{\s T_{\s +}}\!+\! A_{\s T_{\s -}}\right),
\end{align}
where $\chi$ is the quasienergy, see Fig. \ref{re-chi}, while
parameters $\delta_{\s 0}$ and $\delta$ are defined as
\begin{equation}
\label{delta}
\delta_{\s 0} =\frac{1}{2}\left(\delta_{\s e}-\delta_{\s h}\right),~~~ \delta=\frac{1}{2}\left(\delta_{\s e}+\delta_{\s h}\right).
\end{equation}
The quasienergies satisfy the equation
\begin{equation}
\label{quartic}
\chi^2 (\chi^2 - \delta^2 - \Omega_{\s R}^2 ) -
    \delta_{\s 0}^2(\chi^2 - \delta^2) = 0
\end{equation}
with obvious solutions
$\chi =\pm \frac{1}{2}\left[ \left(\delta_{\s 0}+\delta\right)^2+\Omega_{\s R}^2\right]^{1/2}\pm \frac{1}{2}\left[ \left(\delta_{\s 0}-\delta\right)^2+\Omega_{\s R}^2\right]^{1/2}$.
It follows from Eqs. (\ref{driven-system}), (\ref{quartic}) that for large
$\Omega_{\s R} \gg \delta_{\s 0},\delta$, the pair of quasienergies, which approaches $\chi =0$, corresponds to the modes $S$ and $\frac{1}{\sqrt{2}}(T_{\s +} - T_{\s -})$, while the quasienergies  that approach
$\chi =\pm~\Omega_{\s R}$ correspond to the combinations
$\frac{1}{2}\left(T_{\s +}\pm \sqrt{2}T_{\s 0}
+T_{\s -}\right)$, respectively.

\noindent
{\em Driven spin-pair with recombination. }
Including recombination from $S$ requires the analysis of the
full equation for the density matrix, $i\dot{\rho} = [\widehat{H}, \rho] - \frac{i}{2 \tau} \left\{ \Pi_{\s S}, \rho \right\}$,
where $\tau$ is the recombination time, and $\Pi_{\s S}$ is the projector
onto the singlet subspace.
The matrix corresponding to this equation
is $16 \times 16$.
The 16 eigenvalues can be cast in the form
$\chi_{i} - \chi_{j}^{*}$, where
$\chi_i$ and $\chi_j$ satisfy the quartic equation
\begin{equation}
\label{quartic-recombination}
\chi\left( \chi + \frac{i}{\tau}\right) (\chi^2 - \delta^2 - \Omega_{\s R}^2 ) -
    \delta_{\s 0}^2(\chi^2 - \delta^2) = 0,
\end{equation}
which generalizes Eq. (\ref{quartic}) to the pair with decay. For slow recombination, $b_{\s 0}\tau \gg 1$,
the quasienergies acquire small imaginary parts, which can be found perturbatively from Eq. (\ref{quartic-recombination})
\begin{equation}
\label{deltachi}
\delta \chi = -\frac{i}{4 \tau} \left( 1 \pm \frac{\left|\delta_{\s 0}^2 - \delta^2 - \Omega_{\s R}^2\right|}{
 \sqrt{\left( \delta^2 + \delta_{\s 0}^2 + \Omega_{\s R}^2\right)^2 \!\! - 4\delta_{\s 0}^2 \, \delta^2 }} \right).
\end{equation}
Naturally, in the limit $\Omega_{\s R}\rightarrow 0$, Eq. (\ref{deltachi})
yields either $\delta \chi =-i/2\tau$ for $S$ and $T_{\s 0}$ states,
and $\delta \chi =0$ for the trapping states $T_{\s +}$ and $T_{\s -}$.
Less trivial is that at large $\Omega_{\s R} \gg \delta_{\s 0},\delta$ the values $\delta \chi$ again
approach $\delta \chi =-i/2\tau$ and $\delta \chi =0$. The evolution of the imaginary parts of the quasienergies with $\Omega_{\s R}$ is illustrated in Fig. \ref{im-chi}.

\noindent{\em Current at a weak drive.}
Finite $\Omega_{\s R} \ll \delta, \delta_{\s 0} \sim b_{\s 0}$ leads to finite lifetimes of the trapping modes. Expanding Eq. (\ref{deltachi}), we get
\begin{equation}
\label{finite}
\tau_{tr}=\frac{1}{2\vert \chi \vert}=\frac{4\tau\left(\delta^2 -\delta_{\s 0}^2\right)^2}{\Omega_{\s R}^2\delta_{\s 0}^2}.
\end{equation}
Once $\tau_{tr}$ is known, we can employ the simplest
quantitative description of transport\cite{we} based on the model Ref. \onlinecite{Bobbert1}
to express the correction, $\delta I(\Omega_{\s R})$, to the current caused by the ac drive. Within this description, a pair at a given site is first assembled, then undergoes the pair-dynamics and either recombines or gets disassembled depending on which process takes less time, see Fig. \ref{irradiation} (a). These three steps are then repeated, so that the passage of current proceeds in cycles. Then the current associated
with a given pair is equal to  $\frac{1}{\langle t \rangle}$, where ${\langle t \rangle}$ is the average cycle duration.
Importantly, all the initial spin configurations of the pair
have equal probabilities.  For simplicity, it is assumed\cite{we} that, on average, the times of assembly and disassembly are the same $\tau_{\s D}\gg \tau$. This input is
sufficient to derive the following expression for $\delta I(\Omega_{\s R})$
\begin{equation}
\label{deltaI}
\frac{\delta I(\Omega{\s R})}{I(0)}=
\frac{\tau_{tr}^{-1}}{\tau_{tr}^{-1}+2\tau_{\s D}^{-1}}
=\frac{\Omega_{\s R}^2\delta_{\s 0}^2}
{\Omega_{\s R}^2\delta_{\s 0}^2+8(\delta^2-\delta_{\s 0}^2)^2\frac{\tau}{\tau_{\s D}}},
\end{equation}
where $I(0)=\frac{1}{\tau_{\s D}}$. The remaining task is to
average Eq. (\ref{deltaI}) over the distributions of the hyperfine fields, or equivalently, over $\delta$ and $\delta_{\s 0}$. Since we consider a weak drive, this
averaging is greatly simplified. Indeed, the major contributions to the average comes from narrow domains
$\vert \delta-\delta_{\s 0}\vert
\sim \Omega_{\s R}\left(\frac{\tau_{\s D}}{\tau}\right)^{1/2}$ and
$\vert \delta+\delta_{\s 0}\vert
\sim \Omega_{\s R}\left(\frac{\tau_{\s D}}{\tau}\right)^{1/2}$, much narrower than $b_{\s 0}$. On the other hand, these domains are wider than $\Omega_{\s R}$, which justifies the expansion Eq. (\ref{finite}). Replacing the distribution functions of $(\delta+\delta_{\s 0})$ and
$(\delta-\delta_{\s 0})$ by $\frac{1}{\sqrt{\pi}b_{\s 0}}$, we get
\begin{multline}
\label{result}
\frac{\langle \delta I(\Omega_{\s R}) \rangle}{I(0)} =
\frac{\Omega_{\s R}^2}{(2 \pi)^{1/2} b_{\s0}}
\int
\frac{d(\delta - \delta_{\s 0})}{\Omega_{\s R}^2 + \frac{32 \tau}{\tau_{\s D}} (\delta - \delta_{\s 0})^2}  \\
+  \frac{\Omega_{\s R}^2}{(2 \pi)^{1/2} b_{\s0}}
\int
\frac{d(\delta + \delta_{\s 0})}{\Omega_{\s R}^2 + \frac{32 \tau}{\tau_{\s D}} (\delta + \delta_{\s 0})^2}
=
\left(\frac{\pi\tau_{\s D}}{2\tau}\right)^{1/2}
\left(\frac{\Omega_{\s R}}{b_{\s 0}}\right),
\end{multline}
i.e. the radiation-induced correction is {\em linear} in $\Omega_{\s R}$.
To understand this anomalous behavior qualitatively, notice that small $(\delta+\delta_{\s 0})$ and $(\delta-\delta_{\s 0})$
correspond to small $\delta_{e}$ and $\delta_{h}$, respectively.
Therefore, the linear $\delta I(\Omega_{\s R})$ comes from configurations of hyperfine fields in which one
of the pair-partners is on-resonance\cite{Lips,RabiPolymer,Glenn}; this partner
responds strongly to the ac drive. The  ratio $\Omega_{\s R}/b_{\s 0}$ is the portion of such configurations.
The upper boundary of the weak driving domain is set by the condition $\Omega_{\s R} \sqrt{\tau_{\s D}/\tau} \lesssim b_{\s 0}$, which allowed us to replace the distribution
functions of $\delta - \delta_{\s 0}$, $\delta + \delta_{\s 0}$ by a constant. It is
also seen from Eq. (\ref{deltaI}) that for $\Omega_{\s R} \gg b_{\s 0} \sqrt{\tau_{\s D} /\tau}$ that the correction  saturates at $\langle \delta I \rangle/I(0) = 1$.
This saturation applies as long as $T_{\s +}$ and $T_{\s -}$ are the trapping  eigenmodes. As was mentioned above, upon increasing $\Omega_{\s R}$, the trapping eigenmodes evolve into $\frac{1}{2}\left(T_{\s +}\pm \sqrt{2}T_{\s 0}
+T_{\s -}\right)$ and we enter the  strong-driving regime.

\noindent{\em Strong drive.}
Expanding Eq. (\ref{deltachi}) in the limit $\Omega_{\s R}\gg \delta,\delta_{\s 0}$
yields the expression $\tau_{tr}\approx \tau \Omega_{\s R}^2/\delta_{\s 0}^2$
for the lifetime of the trapping eigenmodes.
The same steps that led to Eq. (\ref{deltaI}) give rise to the following {\em negative} correction to the current
\begin{equation}
\label{deltaIprime}
\frac{\delta I(\Omega{\s R}) }{I(0)}=1-\left(\frac{\tau}{\tau_{\s D}}\right)
\frac{\Omega_{\s R}^2}{\delta_{\s 0}^2+\frac{\tau}{\tau_{\s D}}\Omega_{\s R}^2}.
\end{equation}
We see from Eq. (\ref{deltaIprime}) that at $\Omega_{\s R}\gg
\left(\frac{\tau_{\s D}}{\tau}\right)^{
1/2}b_{\s 0}$  the current  is the same as it was in the absence of the ac drive.
This is due to the fact that both in the absence of drive and in this domain the number
of long-living modes is two. The return of $\delta I(\Omega_{\s R})$ to zero takes place
over a parametrically broad interval
$\sqrt{\frac{\tau}{\tau_{\s D}}} < \frac{\Omega_{\s R}}{b_{\s 0}}< \sqrt{\frac{\tau_{\s D}}{\tau}}$. The slope is calculated upon averaging Eq. (\ref{deltaIprime}) over $\delta_{\s 0}$, which again can be carried out after replacing the distribution function
by $\frac{1}{\sqrt{\pi}b_{\s 0}}$ and yields
\begin{equation}
\label{slope}
\frac{1}{I(0)}
 \frac{\partial \langle \delta I \rangle}{\partial \Omega_{\s R}}
  = - \left(\frac{\tau }{\pi \tau_{\s D}}\right)^{1/2} \frac{1}{b_{\s 0}}.
\end{equation}
This result shows a slope which is $\tau_{\s D}/\tau$ times smaller than that given
by Eq. (\ref{result}); this is consistent with the fact that the domain
of the current drop is $\tau_{\s D}/\tau$ times broader than the domain
of current growth.

In fact, the saturation predicted by Eq. (\ref{deltaIprime}) precedes another domain of change of current,
which stems from bifurcation in lifetimes of $S, T_{\s 0}$ modes at large $\Omega_{\s R}$, see Fig \ref{im-chi}. To capture this bifurcation analytically,
notice that for large $\Omega_{\s R}$ Eq. (\ref{deltachi}) predicts
for $\delta \chi = - \frac{i}{2\tau}$ for the $\frac{1}{\sqrt{2}}(T_{\s +} -T_{\s -})$-mode, while the
zero-order value of quasienergy falls off with $\Omega_{\s R}$ as  $\delta_{\s 0}\delta/\Omega_{\s R}$. When $\Omega_{\s R}\gtrsim\delta\delta_{\s 0}\tau$, the correction exceeds the zero-order value and the perturbative treatment becomes inapplicable. Instead, we must make use of the fact that quasienergy is small, which allows us to simplify the quartic equation Eq. (\ref{quartic}) to \begin{equation}
\label{quadratic}
\chi^2 + \frac{i}{\tau} \chi - \frac{\delta_0^2 \delta^2}{\Omega_{\s R}^2} = 0.
\end{equation}
The bifurcation of the lifetimes is revealed in the imaginary
parts of the quasienergies, which are given by
\begin{equation}
\label{quad-soln}
\chi_{\s \pm} = - \frac{i}{2\tau}
\left[1 \pm \sqrt{1 - \frac{4 \delta_0^2 \delta^2 \tau^2}{\Omega_{\s R}^2}}\right],
\end{equation}
see Fig. \ref{im-chi}. For large $\Omega_{\s R}$, solution $\chi_{\s +}\approx - i/\tau$
corresponds to the $S$-mode, while the solution $\chi_{\s -}\approx
-i \delta_0^2 \delta^2 \tau/\Omega_{\s R}^2$ evolves into a long-living mode $\frac{1}{\sqrt{2}}\left(T_{\s +}-T_{\s -}\right)$.
In other words, strong ac drive {\em induces} a third long-living mode which decouples from $S$, and therefore, cannot recombine. At the same time, the decoupling of $S$ from all other triplet 
states makes its lifetime two times shorter than in the absence of drive.
Note, that there is a full formal correspondence between the solutions $\chi_{\s +}$,
$\chi_{\s -}$ and the superradiant and subradiant modes in the Dicke effect\cite{Dicke}.
On the physical level, in the Dicke effect, the subradiant mode acquires a long lifetime due to weak overlap with a photon field, while the long lifetime of the mode $\frac{1}{\sqrt{2}}\left(T_{\s +} - T_{\s -}\right)$ is due to weak overlap with the recombining state $S$.
With trapping by the subradiant mode incorporated, the correction to current takes the form
\begin{equation}
\label{current11}
\frac{\delta I({\Omega_{\s R}})}{I(0)} =- \frac{\Omega_{\s R}^2}{
\left(\delta_{\s 0}^2 \delta^2 \tau \tau_{\s D} + \Omega_{\s R}^2 \right)}.
\end{equation}
It can be seen that the denominator in Eq. (\ref{current11}) defines a narrow domain $\delta_{\s 0} \sim \delta \sim
 \Omega_{\s R}^{1/2}/\left(\tau\tau_{\s D}\right)^{1/4}$,
which yields the major contribution to $\langle \delta I(\Omega_{\s R})\rangle$. Physically, this corresponds to configurations of the hyperfine fields in which {\em both} pair-partners are on-resonance.
This again leads to the linear correction to $\langle \delta I(\Omega_{\s R})\rangle$, which can be
rewritten in dimensionless units as
\begin{equation}
\label{last}
\frac{\langle \delta I (\Omega_{\s R}) \rangle}{I(0)}
=
- \frac{\Omega_{\s R}}{\pi b_0^2 \sqrt{\tau \tau_{\s D}}} \int dx \int dy
\frac{1}{ x^2 y^2 + 1}.
\end{equation}
The double integral in Eq. (\ref{last}) diverges, but only logarithmically, as $\ln \left[b_{\s 0}^2(\tau\tau_{\s D})^{1/2}/
\Omega_{\s R}\right]$.
\begin{figure}[t]
\includegraphics[width=77mm, clip]{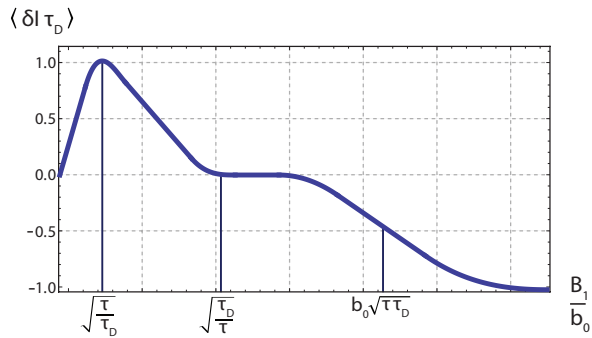}
\caption{(Color online). Schematic dependence of the radiation-induced correction to the
current on the amplitude of the ac drive. Three prominent domains (a), (b), and (c) 
are described by Eqs. (\ref{result}), (\ref{current11}), and (\ref{last}), respectively. }
\label{current}
\end{figure}
In performing the averaging Eq. (\ref{last}) we again replaced the distribution
functions of $\delta$, $\delta_{\s 0}$ by $\frac{1}{\sqrt{\pi} b_{\s 0}}$.
This replacement is justified provided the characteristic $\delta$, $\delta_{\s 0}$
are much smaller than $b_{\s 0}$.  The latter condition is equivalent to the
condition that the argument of the logarithm is big.  We should also check the
validity of the expansion of the square root in Eq. (\ref{quad-soln}). For
characteristic $\delta$, $\delta_{\s 0}$ the combination $\delta^2 \delta_{\s 0}^2 \tau^2 / \Omega_{\s R}^2$ is $\sim \tau / \tau_{\s D} \ll 1$, i.e. the expansion is
valid.
Overall dependence of $\langle \delta I\rangle$ on $\Omega_{\s R}$ exhibiting three prominent domains, Eqs. (\ref{result}), (\ref{deltaIprime}) and (\ref{last}) is sketched in Fig. \ref{current}.

\noindent{\em Discussion.} The prime experimental finding reported in Ref. \onlinecite {Baker}, which motivated the present
paper, is that the current blocking responsible for the OMAR effect\cite{Bobbert1} is effectively lifted under  magnetic-resonance conditions.
We demonstrated that this lifting is a natural consequence of developing of the Rabi oscillations in one of the spin-pair partners.
It is also known\cite{Lips,RabiPolymer,Glenn} that Rabi oscillations in organic 
semiconductors, detected by  pulsed
magnetic resonance techniques, are also dominated by
 pairs in which one partner is on-resonance.
The reason why both effects are due to the same sparse objects
is that these objects are more responsive to the ac-drive than
non-resonant pairs. At the same time, the phase volume of such
pairs is linear in $\Omega_{\s R}$.

Besides the physical picture in the weak-driving domain, we 
also predict that the overall  evolution of current with increasing $B_{\s 1}$ is
much more complex, and involves a maximum followed by a drop and subsequent saturation, see Fig. \ref{current}.
Note that strong deviation from linear dependence of $\delta I$ sets in
already at weak driving fields, $B_{\s 1}\lesssim b_{\s 0}$.
The non-monotonic behavior of current with ac drive is very unusual;  its
experimental verification would be a crucial test of radiation-induced trapping,
which we predict.

Throughout the paper we assumed that the driving frequency exactly
matches the Zeeman splitting $\gamma B_{\s 0}$. In fact, in Ref. \onlinecite {Baker} the
sensitivity of OMAR to the ac drive extended over a sizable interval of applied dc fields
centered at $B_{\s 0}$. It is straightforward to generalize our consideration to a finite
detuning $\Delta = \gamma B_{\s 0}-\omega_{\s 0}$. It enters the theory as a shift of the center of
the gaussian distribution of parameter $\delta$  from $\delta=0$ to $\delta=\Delta$.
Below we simply list the changes in the correction $\delta I$ caused by strong detuning $\Delta \gg \gamma b_{\s 0}$. These changes are different in different domains of the driving field shown in
Fig. \ref{current}. For weak driving the correction $\delta I$ is given by
\begin{equation}
\frac{\delta I (\Omega_{\s R})}{I(0)} =\frac{\Omega_{\s R}^2 b_{\s 0}^2 \tau_{\s D}}{8 \Delta^4 \tau}.
\end{equation}
It emerges upon  neglecting the $\Omega_{\s R}^2$ term in the denominator of Eq. (\ref{deltaI}) and applies in the domain $\Omega_{\s R} \lesssim \Delta$ if $\Delta$ exceeds not only $b_{\s 0}$ but also
$b_{\s 0}\sqrt{\frac{\tau_{\s D}}{\tau}}$.  Then, unlike Fig. \ref{current}, the change $\frac{\delta I}
{I(0)}$ does not reach one. The maximal change is $\sim b_{\s 0}^2\tau_{\s D}/\Delta^2\tau \ll 1$.
Interestingly, the domain (c) in Fig. \ref{current} is affected much weaker by the detuning, $\Delta$.
Instead of Eq. (\ref{last}) we have 
\begin{equation}
\frac{\delta I (\Omega_{\s R})}{I(0)} = - \frac{\Omega_{\s R}}{\Delta b_{\s 0} \sqrt{\pi \tau \tau_{\s D}} },
\end{equation}
which amounts to the suppression of the linear slope by $\Delta/b_{\s 0}$.

\noindent{\em Acknowledgements.}
We are grateful to W. Baker and C. Boehme for piquing
our interest in the subject.
This work was supported by NSF through MRSEC DMR-1121252.

\end{document}